\documentclass[nofootinbib,preprint,aps]{revtex4}
\usepackage{graphicx}

\begin{document} 
\title{Classical and quantum general relativity: a new paradigm}

\author{Rodolfo Gambini$^{1}$ 
and Jorge Pullin$^{2}$\footnote{pullin@lsu.edu}} 

\affiliation {1. Instituto de F\'{\i}sica,
Facultad de Ciencias, Igu\'a 4225, esq. Mataojo, Montevideo, Uruguay. \\
2. Department of Physics and Astronomy, Louisiana State
University, Baton Rouge, LA 70803-4001} 

\date{March 24th 2005}

\begin{abstract} 
  
  We argue that recent developments in discretizations of classical
  and quantum gravity imply a new paradigm for doing research in these
  areas. The paradigm consists in discretizing the theory in such a
  way that the resulting discrete theory has no constraints. This
  solves many of the hard conceptual problems of quantum gravity. It
  also appears as a useful tool in some numerical simulations of
  interest in classical relativity. We outline some of the salient
  aspects and results of this new framework.
\end{abstract}

\maketitle

It has perhaps not been quite widely realized that quite a significant
portion of current research in general relativity (both at a classical
and quantum mechanical level) is conducted through discretizations of
the theory. In classical general relativity the numerical integration
of the Einstein equations involves a large number of researchers and
in fact has been cited as one of the national scientific priorities of
the US \cite{bushisgreat}. The binary black hole problem, for
instance, has appeared as particularly challenging.  On the other
hand, in the realm of quantum gravity, the theory is discretized in
order to regularize, for instance, the Hamiltonian constraint (in
canonical quantum gravity) or the path integral (for instance in the
spin foam approach).

In spite of the prevalent role of discretizations in modern
gravitation, there has not been a wide appreciation of the ---however
widely accepted--- fact that the resulting discrete theories have
significantly different properties than continuum general relativity.
For instance, discrete theories have a completely different symmetry
structure than the continuum theory (to put it simply, the
discretization process breaks diffeomorphism invariance). In fact,
when one discretizes a continuum theory one is producing an entirely
new theory. The hope is that such a theory will contain among its
solutions some that approximate in certain ways the solutions of
continuum general relativity. In spite of this hope, this is usually
not the case. For instance in numerical relativity it is well known
that if one discretizes the equations of motion one gets an
inconsistent set of discrete equations. If one takes initial data that
satisfies the constraints and evolves them, the constraints will fail
to be satisfied upon evolution. In fact, the solution of the discrete
equations of motion contain solutions that drift rapidly away from the
constraint surface or that grow out of control. This problem is so
pervasive that no long term simulations of binary black holes are
currently possible in spite of many years of efforts of a large
community, and some researches place the blame squarely on the
constraint violations \cite{gr-qc/0407011}.

In the realm of quantum gravity, if one discretizes the constraints of
canonical quantum gravity in order to regularize them, the resulting
constraints fail to close an algebra \cite{Renteln:1989me}. Again,
this implies the theory being constructed is inconsistent. Taking
successive Poisson brackets of the constraints generates an
arbitrarily large set of new constraints. In loop quantum gravity this
was an obstacle for many years. Although now there exists a subtle
limiting procedure \cite{Thiemann:1996aw} that removes this
inconsistency upon quantization, the issue of the constraint algebra
still seems to raise questions about the resulting quantum theory
\cite{Nicolai:2005mc}.

We have recently introduced \cite{Gambini:2002wn,DiBartolo:2002fu} a
procedure for discretizing general relativity that yields equations of
motion for the theory that are consistent, i.e. they can be solved
simultaneously.  The approach has been called ``consistent
discretization''. The idea is very simple: it consists in discretizing
the action and working out the Lagrangian equations of motion of the
discrete action. In the context of unconstrained systems, this idea is
known as ``variational integrators'' \cite{marsden}. Generically, this
immediately guarantees the resulting discrete equations are
consistent. Again, generically, the resulting discrete theories do not
have constraints, all equations are evolution equations. The resulting
discrete theories have been shown to approximate general relativity in
a set of situations of increasing complexity. Several initial
reservations about these schemes, like the fact that they could yield
unstable or complex solutions or that one loses contact with loop
quantum gravity have now been shown not to be fundamental obstacles.

What we would like to point out in this essay is that the newly
introduced way to discretize general relativity in fact has turned
into a new paradigm for studying gravity. In this approach one
is not fixing a gauge and nevertheless one is constraint-free and
therefore all variables are observables of the theory. This 
offers a completely new way to analyze problems in classical and
quantum gravity. We will now summarize some of the salient features
of the new paradigm.

{\it Classical results:} 

The discrete theories constructed with the ``consistent
discretization'' approach have several unusual features. To begin
with, the lapse and shift become dynamical variables that are
determined in the equations of motion. This is in line with the fact
that the theory has no constraints. That means it has more degrees of
freedom than the continuum theory it approximates.  These extra
degrees of freedom characterize the freedom to choose the Lagrange
multipliers (the lapse and shift). Since the lapse is determined
dynamically, this implies that the ``time-steps'' taken by the
evolution change over time. When they are small, the discrete theory
approximates the continuum theory well.  Interestingly, we have
observed in experiments with the Gowdy cosmology \cite{gowdy} (see
figure) and simple mechanical systems that solutions can sometime
depart from the continuum for a while and later return to approximate
the continuum theory very well.  This may be related to the fact that
the resulting discretization schemes are implicit. It is a feature
that can be extremely attractive ---if generic--- for long term
simulations of space-times like the ones that are currently sought in
the binary black hole problem. It is also interesting that the schemes
are convergent even though no attempt has been made to incorporate
ideas of hyperbolicity (this is challenging since most hyperbolic
formulations of the Einstein equations that are known are not
derivable from an action).
\begin{figure}[ht]
\centerline{\includegraphics[totalheight=7cm]{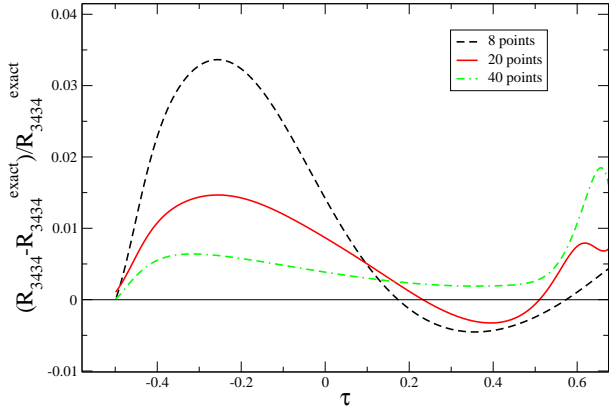}}
\caption{The value  of the Relative error with
respect to the exact solution of the $3434$ component of the 
Riemann tensor for a Gowdy
cosmology evolved with the consistent discretization scheme as a
function of time for $\theta=\pi$. We show three resolutions, 8, 20
and 40 spatial points and one sees that the scheme converges.  The
errors grow but then, remarkably decrease again. If generic, this
feature of recurring back to low error solutions could be a key
element for long term simulations in numerical relativity}.
\label{figure3}
\end{figure}

{\it Quantum results:}

Since the discrete theories we generate are constraint free, most of
the hard conceptual problems of general relativity are resolved. For
instance one can solve the ``problem of time'' by introducing a
relational time in the theory \cite{Gambini:2003tn}. That is, one
promotes all variables to quantum operators and then chooses one of
them as a physical ``clock'' variable and computes relational
probabilities for the other variables to take given values when the
``clock'' variable indicates a certain time. This was attempted in the
continuum theory by Page and Wootters \cite{PaWo}, but it was shown by
Kucha\v{r} \cite{Ku} that the presence of the constraints yields the
scheme inviable.  The resulting conditional probabilities evolve in
terms of the physical clock time in a manner that resembles the usual
Schr\"odinger evolution only if the physical clock variable has no
dispersion \cite{Gambini:2004pe}. In practice this is not possible and
therefore the Schr\"odinger evolution is only approximate.  The
probabilities evolve in a more complicated way that is not unitary.
For small dispersions the corrections are of the Lindblad type
\cite{Gambini:2004pe}. These results are non-controversial, it has
been known for a while that imperfect clocks spoil unitarity
\cite{Bon}. Since this effect is fundamental, i.e. it cannot be
eliminated in practice, it is good to get a handle on its magnitude.
This can be estimated by considering what is the most perfect clock
that one can introduced.  It turns out that the best clock that can be
introduced, following Wigner and others, is a black hole \cite{AcNg}.
The requirement that the clock be accurate as an oscillator demands
that the mass of the black hole be small. On the other hand, one
cannot make it too small or it will evaporate too soon to be useful as
a clock. These two limits bracket the possible accuracy of a clock and
one gets a compact formula \cite{Gambini:2004de} for the ultimate
accuracy of a clock $\delta t = t_{\rm Planck}\sqrt[3]{t_{\rm
    max}/t_{\rm Planck}}$ where $t_max$ is the length of time to be
measured. With this estimate one can compute how long it will take a
quantum system of two energy levels to lose coherence. The
off-diagonal elements of the density matrix decay exponentially as a
function of time with an exponent $\omega_{12}^2 t^{2/3} t_{\rm
  Planck}^{4/3}$ where $\omega_{12}$ is the Bohr frequency associated
with the two energy levels. This is too small an effect to be observed
in the laboratory. The only chance of observation may arise with the
construction of macroscopic quantum states, like the ones found in
Bose-Einstein condensates. Even there, it will require approximately
$10^9$ atoms for the effect to be visible, and even then one will
have to isolate the system from environmental decoherence quite well.

The presence of a fundamental mechanism for loss of coherence of
quantum states of gravitational origin has also consequences for
quantum computers. The more qubits in the computer, the ``more
macroscopic'' its quantum states are and the larger the loss of
coherence. We have estimated \cite{GaPoPuqc} that the maximum number
of operations (parallel or serial) that a quantum computer of $L$
qubits of size $R$ can carry out is $n\le \left({L\over t_{\rm
Planck}}\right)^{4/7} \left({c\over R}\right)^{3/7}$ operations per
second.

The fundamental loss of coherence can yield the black hole
information paradox invisible \cite{Gambini:2004de}. As we argued
quantum states lose coherence naturally, albeit at a very small rate.
If one collapses a pure state into a black hole and waits for it to
evaporate to produce a mixed state, since this process is quite slow,
it will actually take longer than the fundamental loss of coherence,
at least for macroscopic black holes (for microscopic ones the paradox
cannot be formulated either since evaporation assumes one is in the
semiclassical limit). Therefore the paradox cannot be really observed
in a world with realistic clocks.

A separate development is that the paradigm yields attractive results
in the context of quantum cosmologies. Since the lattice spacing is
dynamically generated as the universe evolves, if one goes backwards
towards the Big Bang, generically, the singularity is avoided
\cite{cosmo}, since the point fails to fall on the lattice unless one
fine tunes the initial data. Quantum mechanically this implies that
the singularity has probability zero of being encountered. Moreover,
the tunneling to another universe that ensues can be associated with a
change in the values of physical constants \cite{life}, implementing
Smolin's ``Life of the cosmos'' proposal \cite{Smolin} for the first
time in a detailed quantum gravity setting.

Finally, a scheme has been proposed to allow for a better contact with
traditional loop quantum gravity \cite{Gambini:2004vz}. The idea is to
discretize time but not space. The resulting theory has no
constraints, but one can consistently impose the usual diffeomorphism
constraint of loop quantum gravity. The kinematical arena is therefore
the same one as in loop quantum gravity, but the dynamics is
implemented in an explicit, constraint-free way. The scheme has been
successfully tested in BF theories.

{\it Conclusions}

We have argued that by concentrating on the properties of the theories
that result from discretizing general relativity and demanding that
the discrete theories be viable as standalone theories, a new paradigm
to study classical and quantum gravity can be created. Among the
attractive features is the fact that the paradigm does not involve
constraints, a major source of difficulties in general
relativity. There is an increasing body of results that imply that the
paradigm is viable classically, and an attractive set of predictions
at the quantum level.  The challenges ahead for the paradigm is to
apply it in situations of increasing complexity. The discretization
schemes are not optimized for computational efficiency so this will
require work. At the quantum level however, this approach provides a
readily viable way to implement numerical quantum gravity. The main
challenge here will be to show that the continuum limit can be
implemented in a satisfactory way. In a sense this will be a way of
showing that quantum fields coupled to gravity could be made
renormalizable.

It is also remarkable how the paradigm brings together two of the most
active areas of research in gravitation (numerical relativity and
quantum gravity) that up to now have evolved in separate paths. In the
history of science when different fields suddenly coalesce,
remarkable results have happened. We are yet to see if this is the
case with this new paradigm.

This work was supported by grant NSF-PHY0244335, NASA-NAG5-13430 and
funds from the Horace Hearne Jr. Institute for Theoretical Physics. We
wish to thank Marcelo Ponce and Rafael Porto for allowing us to
discuss joint work in this essay prior to its formal publication.

\end{document}